\documentclass[conference]{IEEEtran}
\IEEEoverridecommandlockouts
\usepackage[numbers,sort&compress]{natbib}
\usepackage{amsmath,amssymb,amsfonts}
\usepackage{algorithmic}
\usepackage{graphicx}
\usepackage{textcomp}
\usepackage{xcolor}
\usepackage{makecell}
\usepackage{multirow}
\usepackage{changes}

\usepackage{flushend}
\usepackage{lipsum}
\usepackage[bottom]{footmisc}

\makeatletter
\newcommand{\linebreakand}{%
\end{@IEEEauthorhalign}
\hfil\mbox{}\par
\mbox{}\hfill\begin{@IEEEauthorhalign}
}
\makeatother
\makeatletter
\renewcommand\@makefntext[1]{%
	\noindent\makebox[0pt][r]{\@makefnmark}#1}
\makeatother

\usepackage{hyperref}
\hypersetup{pdfborder={0 0 0}}
\def\BibTeX{{\rm B\kern-.05em{\sc i\kern-.025em b}\kern-.08em
    T\kern-.1667em\lower.7ex\hbox{E}\kern-.125emX}}
\begin{document}

\title{Arrhythmia Classifier Based on Ultra-Lightweight Binary Neural Network
}

\author{\IEEEauthorblockN{Ninghao Pu}
\IEEEauthorblockA{\textit{School of Electronic Science and}\\ \textit{Engineering} \\
\textit{Southeast University}\\
Nanjing, China\\
213192652@seu.edu.cn}
\thanks{979-8-3503-2138-8/23/\$31.00 ©2023 IEEE}
	
\and
\IEEEauthorblockN{Zhongxing Wu}
\IEEEauthorblockA{\textit{School of Electronic Science and}\\ \textit{Engineering} \\
\textit{Southeast University}\\
Nanjing, China \\
220216191@seu.edu.cn}
\and
\IEEEauthorblockN{Ao Wang}
\IEEEauthorblockA{\textit{School of Electronic Science and}\\ \textit{Engineering} \\
\textit{Southeast University}\\
Nanjing, China \\
Ao0323@outlook.com}
\and

\linebreakand

\IEEEauthorblockN{Hanshi Sun}
\IEEEauthorblockA{ \textit{School of Electronic Science and}\\ \textit{Engineering} \\
\textit{Southeast University}\\
Nanjing, China \\
preminstrel@gmail.com}
\and

\IEEEauthorblockN{Zijin Liu}
\IEEEauthorblockA{\textit{School of Electronic Science and}\\ \textit{Engineering} \\
\textit{Southeast University}\\
Nanjing, China\\
yzliuzijin@163.com}
\and

\IEEEauthorblockN{Hao Liu}
\IEEEauthorblockA{\textit{School of Electronic Science and}\\ \textit{Engineering} \\
\textit{Southeast University}\\
Nanjing, China\\
nicky\_lh@seu.edu.cn}
}

\maketitle

\begin{abstract}
Reasonably and effectively monitoring arrhythmias through ECG signals has significant implications for human health. With the development of deep learning, numerous ECG classification algorithms based on deep learning have emerged. However, most existing algorithms trade off high accuracy for complex models, resulting in high storage usage and power consumption. This also inevitably increases the difficulty of implementation on wearable Artificial Intelligence-of-Things (AIoT) devices with limited resources. In this study, we proposed a universally applicable ultra-lightweight binary neural network(BNN) that is capable of 5-class and 17-class arrhythmia classification based on ECG signals. Our BNN  achieves 96.90\% (full precision 97.09\%) and 97.50\% (full precision 98.00\%) accuracy for 5-class and 17-class classification, respectively, with state-of-the-art storage usage (3.76 KB and 4.45 KB). Compared to other binarization works, our approach excels in supporting two multi-classification modes while achieving the smallest known storage space. Moreover, our model achieves optimal accuracy in 17-class classification and boasts an elegantly simple network architecture. The algorithm we use is optimized specifically for hardware implementation. Our research showcases the potential of lightweight deep learning models in the healthcare industry, specifically in wearable medical devices, which hold great promise for improving patient outcomes and quality of life. Code is available on: \href{https://github.com/xpww/ECG_BNN_Net}{https://github.com/xpww/ECG\_BNN\_Net}.
\end{abstract}

\section{Introduction}
Cardiovascular diseases (CVDs) are the primary cause of death worldwide, killing more people than any other cause\cite{b1}. Arrhythmia, a type of CVDs, is considered to be responsible for most cases of sudden cardiac death that occur each year\cite{b2}. Since the global outbreak of the COVID-19 virus in 2019, arrhythmia has become a common manifestation of cardiovascular disease in COVID-19 patients, and a series of clinical case studies of hospitalised patients in China reported palpitation as the initial symptom in 7.3\% and arrhythmia in 16.7\%\cite{b3}. Electrocardiography currently plays an irreplaceable role in the clinical diagnosis of heart disease and can be used to help medical staff identify the specific type of arrhythmia\cite{b4}, \cite{b5}.

As wearable AIoT technology continues to advance, there is growing interest in implementing ECG classification algorithms on wearable devices. This would enable real-time monitoring of heart health status without the need to visit a hospital. Low-power and low-computational-cost ECG monitoring algorithms are becoming increasingly popular\cite{b6, b7, b8}, but most of them still rely on multi-bit fixed-point computation, which is a significant contributor to computational resource consumption and storage costs. Moreover, the majority of studies also design complex algorithms for pre-processing ECG signals.

To address the aforementioned issues and further reduce hardware costs while conserving computational resources, we employed binarization algorithm. This technique converts the weights and activations of the neural network to $\left\{+1, -1\right\}$, which can be stored in hardware as 1-bit $\left\{0, 1\right\}$ values, effectively reducing storage costs by 32x. Additionally, the convolution operation is converted to bit-wise XNOR-POPCOUNT, aligning seamlessly with the underlying hardware arithmetic logic\cite{b9}. The primary contributions of this article are as follows:
\begin{itemize}
	\item Firstly, a universal hardware-friendly network has been proposed. By quantizing the activations and weights to $\left\{+1, -1\right\}$, fixing the network structure parameters, operator fusion, and replacing the softmax layer, we have created a binary neural networks (BNN) that is easy to program for hardware and can switch between 5-class and 17-class classification arrhythmia monitoring modes.
	\item Secondly, our structure eliminates the need for fully connected(FC) layers. FC layers are almost universally employed in neural networks, their removal resulting in a significant reduction in both storage requirements and computational complexity of the model.
	\item Finally, our 5-class and 17-class arrhythmia classification models achieved remarkable storage compression rates of 29.71x and 29.85x, respectively, with storage usages of 3.76KB and 4.45KB, while maintaining high accuracies of 96.90\% and 97.50\%, respectively. The storage consumption of the model and the accuracy of the 17-class arrhythmia classification model have reached the state-of-the-art level among known lightweight models.
\end{itemize}
\section{Related Work}
Pattern recognition is widely used for monitoring arrhythmias. Prior to classification, signal preprocessing techniques such as filtering and wavelet transformation are applied to extract features\cite{b10,b11,b12}. While this method can achieve satisfactory accuracy, implementing feature extraction algorithms such as filtering and wavelet transformation on wearable devices requires additional resources. Additionally, the high dependence of algorithm performance on the extracted features also greatly reduces the model's generalization ability\cite{b13}, \cite{b14}.

In recent years, with the development of computing power, the computation cost and power consumption of neural networks have increased rapidly. One of the optimal solutions to address these issues is neural network quantization. The most commonly used method of quantization is converting floating-point numbers to 8-bit fixed-point\cite{b15}. Further advances in quantization have resulted in 4-bit quantization, which has only led to approximately 2\% loss in accuracy for AlexNet and VGG16\cite{b16}. The most extreme form of quantization is BNN, which convert both weights and activations to $\left\{+1, -1\right\}$\cite{b9}.

With the development of lightweight models and the focus on implementing ECG classification algorithms in wearable devices, an increasing number of lightweight neural networks have emerged. In \cite{b17}, a hierarchical quantization algorithm based on a greedy algorithm was used to classify ECG with varying bit-widths for different network layers due to their importance. In \cite{b18}, the activations and weights of the ECG arrhythmia classification CNN were both quantized to 8 bits. Farag et al.\cite{b19} proposed a lightweight independent short-time Fourier transform (STFT) convolutional neural network model and quantized it to 8 bits. In \cite{b20}, an adaptive sensor node architecture was designed, and the proposed CNN model was quantized to 8 bits. Sivapalan et al.\cite{b21} converted floating-point operations to 16-bit and 32-bit fixed-point operations for implementation on an embedded platform. Sun el al.\cite{b22} used the Adaptive Loss-aware Quantization (ALQ) algorithm to classify 17-class arrhythmias. Janveja et al.\cite{b6} proposed a low power work and also quantized the model.

In the existing body of research, there are relatively few studies that focus on quantizing the weights and activations of neural networks to $\left\{+1, -1\right\}$ for ECG classification. Wu et al.\cite{b23} designed a full-precision CNN with 9 layers of 1D convolution and 1 fully connected layer, which is then binarized. However, they only utilized F1-score as the evaluation metric for their model, without providing other relevant information such as accuracy, which makes it difficult to compare their results with other models in a comprehensive manner. Wong et al.\cite{b24} designed a BNN to classify ECG signals into V-class and non-V-class, implemented on an FPGA. Their model demonstrates exceptional energy efficiency and superior accuracy. However, we observed that the proposed network's fully connected layer constitutes more than 97\% of the total parameters, and dividing ECG data into only two categories falls short of meeting the needs of both doctors and patients. Wang et al.\cite{b25} proposed a 15-layer CNN to classify ECG and binarized the model. The accuracy of their model requires improvement, and optimization of the batch normalization layer has not been taken into account. During batch normalization computation, floating-point operations are still necessary. Additionally, the parameters in the BNN architecture are not uniform, which poses considerable challenges for the hardware design. Yun et al.\cite{b26} binarized the CNN+LSTM model used for classifying psychological health based on ECG, greatly reducing inference latency and energy consumption. Their model achieves 87\% accuracy with a memory footprint of 47KB, indicating ample room for further improvement.

\section{Methodology}
This section will introduce our network architecture and the binarization method. Our approach has systematically addressed the deficiencies in the relevant literature.

\subsection{Convolutional Neural Network (CNN) Architecture}
To simplify software and hardware programming and facilitate future hardware design, the main structure of the CNN is composed of numerous blocks with standardized parameters such as convolution kernel size, pooling size, pooling stride, etc. To minimize the number of weights in the CNN, we creatively removed the fully connected layers, which are commonly seen in CNN classification networks. This also eliminates the need to separately consider Processing Elements (PE) for FC layers. To reduce the dimensionality of the convolutional output for subsequent classification, we implemented GSP\cite{b27}, which eliminates the division operation in GAP and renders the entire CNN architecture almost free of division operations. To avoid the exponential operation caused by using the softmax function for classification, we replaced it with a comparator, with the index of the maximum value serving as the classification result.

By setting the number of channels in the output of the last block to the desired number of classes and combining it with GSP's dimension reduction, we can simply compare the output of the last dimension of GSP to obtain the specific type of arrhythmia.

Through the aforementioned efforts, we have addressed the following issues in works \cite{b23,b24,b25,b26}: insufficient number of categories and limited model flexibility, substantial parameter consumption resulting from the presence of fully connected layers, and hardware design difficulties stemming from the intricate model architecture. 

\subsection{Method of Binarization}
Floating-point operations are costly and not supported by most embedded hardware. Binary neural network, on the other hand, utilizes XNOR and POPCOUNT operations to convert the floating-point MAC operations of convolutional and fully connected layers into bit-wise operations. This maximizes the reduction in computational complexity and memory usage of weights. We use \eqref{sign(x)} to transform the weights and activation values into $\left\{+1, -1\right\}$.In \eqref{sign(x)}, $x$ represents a floating-point number, and $x^b$ denotes $x$ after binarization.

\begin{equation}
	{x}^{b}=Sign(x)=\left\{
	\begin{aligned}
		& +1, & &x \geq 0& \\
		& -1, & &otherwise&
	\end{aligned}
	\right.
	\label{sign(x)}
\end{equation}

During network training, network are performed using $x^b$. However, when updating weights using loss values, the floating-point values of the weights prior to being converted by the Sign function will be updated. The gradient of the Sign function is infinite at 0 and 0 elsewhere, making it impossible to perform gradient calculations based on the chain rule. We adopt Ding et al.'s Information-Enhanced Estimator(IEE)\cite{b38} to approximate the derivative of the Sign function.

\subsection{Optimizing Batch Normalization Through Operator Fusion}

The aforementioned methodology enabled the successful training of BNN. However, conventional network architectures entail additional operations such as activation functions, batch normalization, and pooling. Prior works \cite{b23}, \cite{b25}, \cite{b26} have utilized batch normalization, but without any additional processing optimizations, this operation will introduce additional floating-point computations, consume hardware resources, and increase computational latency.

Note that PReLU, batch normalization, and Sign function can be mathematically fused into a threshold function, we introduce this optimization method in the ECG classification network to address the aforementioned issue. We use PReLU instead of the more common ReLU because it can improve the representation of BNN\cite{b33}. Below are the specific computational procedures.

\begin{equation}
	f(x)= \begin{cases}x & \text { if } x \geq 0 \\ a x & \text { otherwise }\end{cases}
	\label{prelu}
\end{equation}
\begin{equation}
	y=\frac{x-\mu}{\sqrt{\sigma^2+\varepsilon}} \gamma+\beta 
	\label{bn}
\end{equation}

Equation \eqref{prelu} and \eqref{bn} correspond to PReLU and batch normalization respectively. $a$ is a learnable parameter that represents the slope of the left-hand side line on the y-axis. $\mu$ and $\sigma$ are respectively the mean and variance of the input data calculated during training, which are saved in the model after training. $\varepsilon$ is a very small value set to prevent calculation errors when $\sigma$ equals 0. $\gamma$ and $\beta$ are trainable parameters of batch normalization, which are updated continuously during training through backpropagation of gradients. If batch normalization is placed after PReLU in the network, we can merge them into \eqref{prelu_bn}.
\begin{equation}
	y= \begin{cases}k x+b & \text { if } x \geq 0 \\ a k x+b & \text { otherwise }\end{cases} 
	\label{prelu_bn}
\end{equation}
Here are the values for k and b:
\begin{gather}
	k=\frac{\gamma}{\sqrt{\sigma^2+\varepsilon}} \\
	b=\beta-\frac{\mu \gamma}{\sqrt{\sigma^2+\varepsilon}}
\end{gather}
Continuing on, we introduce a function called $Sign_{\delta}$, where $\delta$ is referred to as the threshold of this function.
\begin{equation}
	{y^b}=Sign_{\delta}(x)=\left\{
	\begin{aligned}
		& +1, & &x \geq \delta& \\
		& -1, & &otherwise&
	\end{aligned}
	\right.
	\label{sign_delta(x)}
\end{equation}
If the function following the batch normalization is Sign function, then \eqref{prelu_bn} can be further merged with sign to obtain \eqref{pre_bn_sign}, which $\delta_{+}$ is $\frac{-b}{k}$ and $\delta_{-}$ is $\frac{-b}{ak}$.
\begin{equation}
	y= \begin{cases}{Sign}_{\delta_{+}}(x) & \text { if } x \geq 0 \\ {Sign}_{\delta_{-}}(x) & \text { otherwise }\end{cases}
	\label{pre_bn_sign}
\end{equation}

Therefore, it is crucial to connect the operators PReLU, BatchNorm, and Sign in sequence to enable operator fusion. While placing Maxpool after Sign can transform the comparator into an OR gate and facilitate fast comparison of 1-bit numbers, early experiments have revealed that this approach leads to significant loss of precision. Therefore, Maxpool can only be placed after the Conv operator. Ultimately, the order of operators within a block can be established as follows: \textbf{Conv, MaxPool, PReLU, and BatchNorm}. The reason why the Sign function was not listed is because it is integrated within the Convolution layer of the next block. Specifically, the binarization of the previous layer's activations and the convolution layer's own weights are both carried out within the convolution layer itself.

\section{Experiment}
In this section, we first introduce the dataset and evaluation metrics used in the experiment. Then, we present the specific parameters of the BNN structure. Next, we present the details and data of the experiment. Finally, we compare our experimental results with the state-of-the-art lightweight models developed in recent years.
\subsection{Dateset}
The network in this experiment is used for both 17-classification and 5-classification of ECG data. All data used in the experiment are sourced from the MIT-BIH Arrhythmia database of PhysioNet\cite{b28}.

The 17-class dataset is the same as Pławiak et al.\cite{b29}, which includes normal sinus rhythm, pacemaker rhythm, and 15 types of heart failure (with at least 10 signal segments collected for each type). Pławiak et al. randomly selected 1000 10-second ECG signal segments for analysis. 

The 5-class dataset contains 7740 10-second segments, and the classification follows the AAMI EC57 standard \cite{b30}, dividing all data into five categories: ventricular ectopic beat (V), normal or bundle branch block beat (N), supraventricular ectopic beat (S), fusion beat (F), and unknown beat (Q). 

Regardless of whether it is the 5-class or 17-class data, we used a signal from one lead (MLII), recorded at 360HZ with 3600 samples in 10 seconds.
\subsection{Evaluation Metrics}
The utilization of diverse and rational evaluation metrics allows for a comprehensive assessment of network performance from various perspectives. To holistically evaluate model performance, we employ a variety of indicators including accuracy (ACC), sensitivity (SEN), specificity (SPE), precision (PRE), and F1-score.

\subsection{Specific Parameters of the Network Structure}
After conducting extensive experiments, we have determined that the optimal number of blocks for BNN is 6. The structures of the 5-class model and the 17-class classification model are almost identical, with the only difference being the output channels of block 6, which correspond to the number of categories for the corresponding arrhythmia. Therefore, by modifying only the final output channel and replacing the network weights, the classifier can switch between 5-class and 17-class classification, endowing the model with universality. Table~\ref{CNN Str Param} presents the specific structural parameters of the proposed BNN. Based on this structure, our model can achieve an accuracy of 97.09\% for 5-class classification and 98.00\% for 17-class classification at full precision.

\begin{table}[htbp]
	\caption{Specific Structural Parameters of BNN}
	\begin{center}
		\setlength{\tabcolsep}{0.5mm}{
			\begin{tabular}{c|c|c|c|c|c}
				\hline 
				&  \scriptsize In Channel & \scriptsize Out Channel & \scriptsize \makecell{Conv Padding \\ and Value$^{\mathrm{a}}$} & \scriptsize \makecell{Conv Kernel\\Size and Stride} & \scriptsize \makecell{Max Pool\\Size and Stride}\\
				\hline
				Block1 & 1 & 8 & 5, 1 & 7,2 & 7,2 \\
				Block2 & 8 & 16 & 5, 1 & 7,1 & 7,2 \\
				Block3 & 16 & 32 & 5, 1 & 7,1 & 7,2\\
				Block4 & 32 & 32 & 5, 1 & 7,1 & 7,2\\
				Block5 & 32 & 64 & 5, 1 & 7,1 & 7,2\\
				Block6 & 64 & 5 or  17 & 5, 1 & 7,1 & 7,2\\
				\hline
				\multicolumn{6}{p{1\linewidth}}{$^{\mathrm{a}}$In our experiments conducted in full precision (i.e., using float32 inference), the padding value was set to 0. On the other hand, in the BNN experiments, the padding value was set to 1.}
		\end{tabular}}
		\label{CNN Str Param}
	\end{center}
\end{table}

\begin{figure}[htbp]
	\centerline{
		\includegraphics[width=0.5\textwidth]{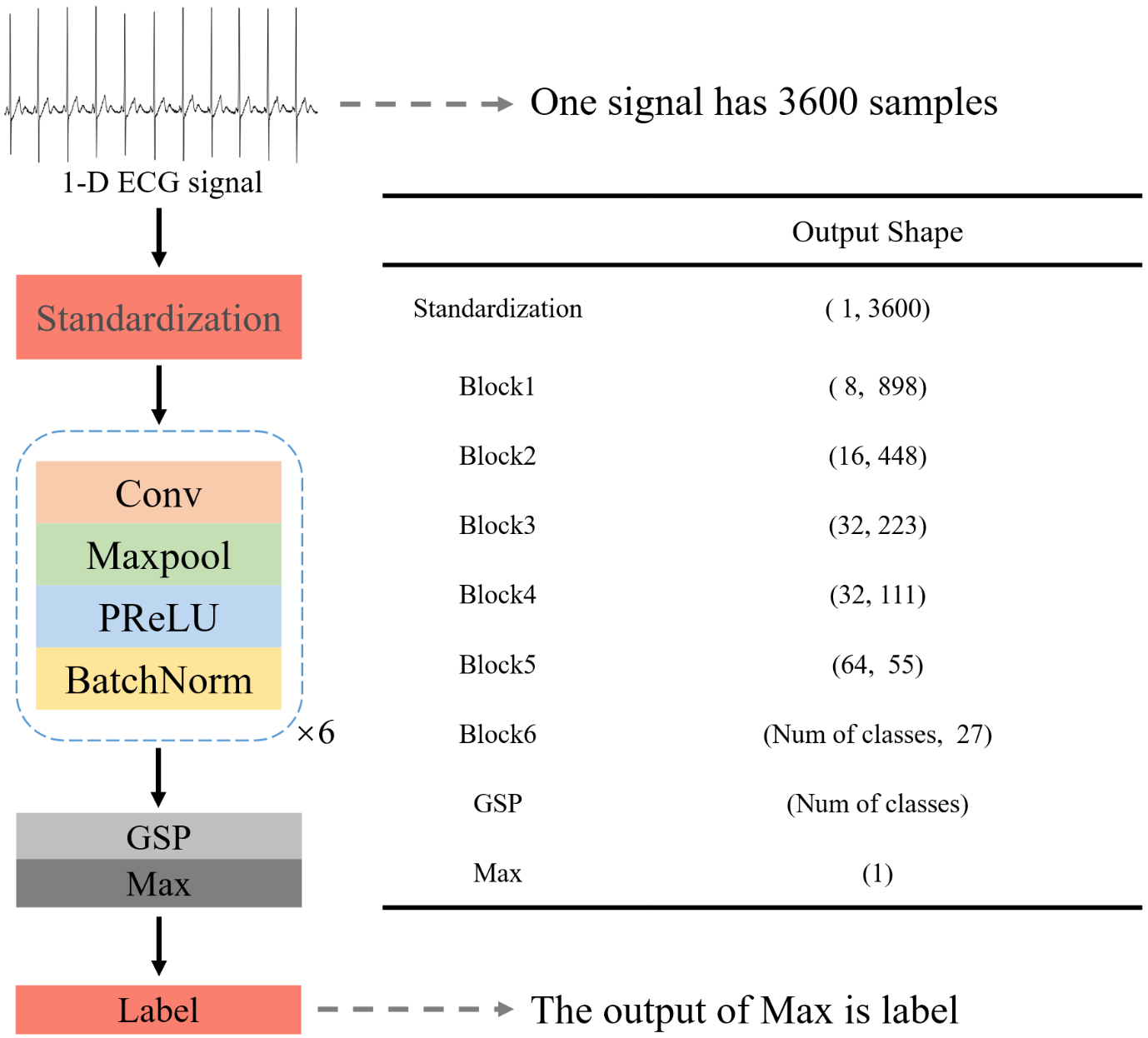}}
	\caption{Binary Neural Network (BNN) structure and the dimensional changes of the data}
	\label{fig1}
\end{figure}

\subsection{Experimentation of Binarization Methods}
For the 5-class classification of arrhythmia, our experimental setup utilized a batch size of 512 and a learning rate of 0.02. For the 17-class classification of arrhythmia, we employed a batch size of 64 and a learning rate of 0.002. All datasets were partitioned such that 80\% was allocated to the training set and 20\% to the testing set, and the models were trained for 1000 epochs.

We tried several popular BNN optimization methods and found that IEE\cite{b38} achieved the best accuracy of 96.90\% in the 5-class classification model, which is only 0.19\% lower than the full precision model. As for the 17-class classification BNN model of arrhythmia, the accuracy always fluctuates between 96.50\% and 97.50\%, and no method can further improve its accuracy. We refer to the models with accuracies of \textbf{96.90\%} and \textbf{97.50\%} as the BP (Better Performance) models.

In the end, we also quantize the ECG signal to $\left\{+1, -1\right\}$ and treat it as an "extreme denoising" technique. Under this circumstance, the first convolution layer can be replaced with XNOR+POPCOUNT. We refer to the model processed in this way as an LP (Low Power) model. Experimental results demonstrate that even under such stringent conditions, our ECG 5-class and 17-class classification models can achieve accuracies of \textbf{91.50\%} and \textbf{92.00\%}, respectively.

The storage space required for the 5-class classification model is \textbf{3.76KB}, achieving a compression ratio of 29.71x. The corresponding arrhythmia 17-class classification model requires a total storage of \textbf{4.45KB}. This results in a memory compression of 29.85x.

\begin{figure}[htbp]
	\centerline{
		\includegraphics[width=0.45\textwidth]{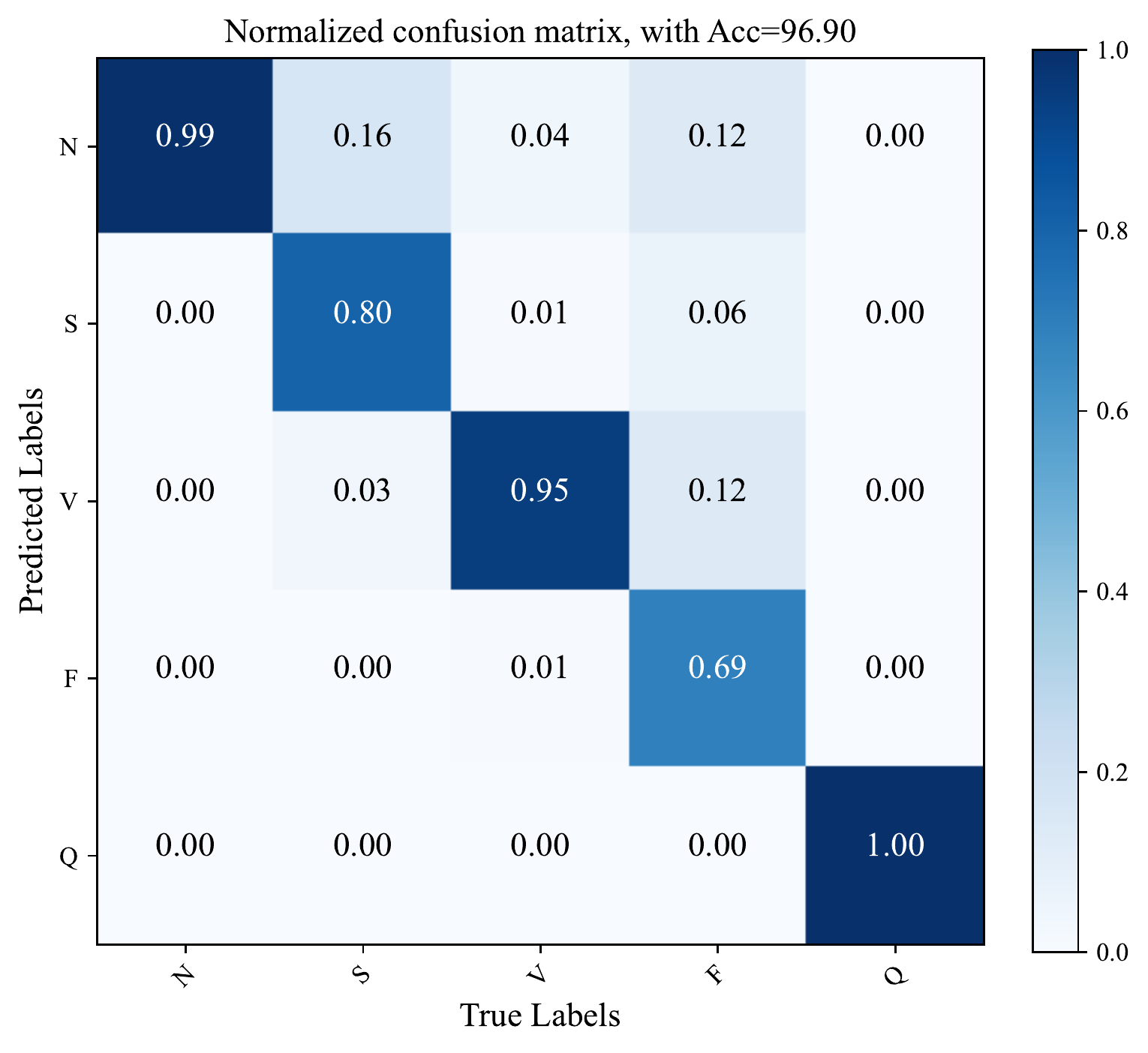}}
	\caption{Confusion matrix for the 5-class ECG classification model}
	\label{Confusion-matrix-5}
\end{figure}

\begin{figure}[htbp]
	\centerline{
		\includegraphics[width=0.5\textwidth]{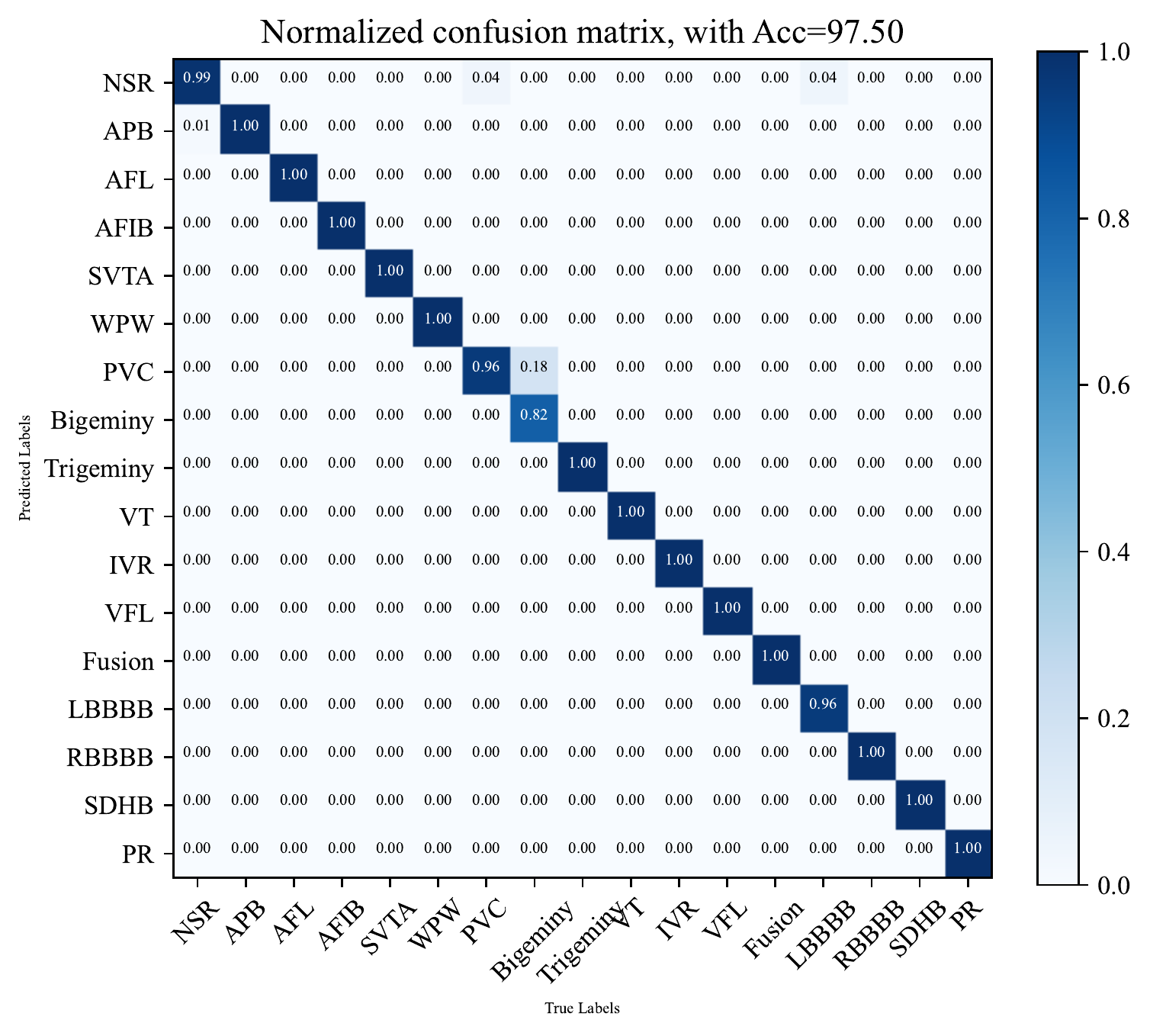}}
	\caption{Confusion matrix for the 17-class ECG classification model}
	\label{Confusion-matrix-17}
\end{figure}

 Due to space limitations, we only provide the confusion matrix of the BP model in Fig.~\ref{Confusion-matrix-5} and Fig.~\ref{Confusion-matrix-17}.

\begin{table*}[htbp]
	\caption{Comparison with Other State-of-the-Art Models}
	\begin{center}
		
		\begin{tabular}{|c|c|c|c|c|c|c|c|c|c|c|}
			\hline
			Work & Features & Methods & Bitwidth(W/A) & Diseases & ACC & SEN & SPE & PRE & F1 & Memory(KB)				 \\
			\hline
			Scrugil et al.\cite{b20} & \makecell{Filtering\\Peak detection} & 2D-CNN & 8/8 & NSVFQ & 96.98 & 95.35 & - & 85.17 & 91.12 & -\\
			\hline
			Sivapalan et al.\cite{b21} & \makecell{SMOTE \\ Denoising \\ Filtering \\ Resampling} & \makecell{LSTM\\MLP} & \makecell{Pre-process \\ and \\ Feature- \\ extraction:\\ 16/16 \\ ANNet:\\32/32} & \makecell{Normal and \\ Abnormal} & 97 & 87 & 98 & 87 & 87 & -\\
			\hline
			Sun et al.\cite{b22} & Normalize & 1D-CNN & Mix & 17-classes & 95.84 & 94.19 & 99.67 & - & - & 13.54 \\
			\hline	
			Wu et al.\cite{b23} & \makecell{Bucket padding \\ Normalize} & 1D-BNN & 1/1 & 4-classes & - & - & - & - & 86.8 & 266.24 \\
			\hline			
			Wong et al.\cite{b24} & Resampling & 2D-BNN & 1/1 & \makecell{V and \\non-V beats} & 97.3 & 91.3 & 98.1 & 86.7 & 88.9 & 8.02\\
			\hline
			Wang et al.\cite{b25} & Normalize & 1D-BNN & 1/1 & NSVFQ & 95.67 & 94.8 & - & 96.2 & - & 10.62\\
			\hline
			Tuncer et al.\cite{b43} & \makecell{DW-CMT + TCP \\ with 128 features} & kNN & 32/32 & 17-classes & 96.60 & 98.51 & - & 95.18 & 96.69 & -\\
			\hline
			\multirow{2}{*}{Ours} & \multirow{2}{*}{Standardize} & \multirow{2}{*}{1D-BNN} & \multirow{2}{*}{1/1} & NSVFQ & \textbf{96.9} & 88.7 & \textbf{98.5} & 91.3 & 89.9 & \textbf{3.76}\\
			\cline{5-11}
			&  &  & & 17-classes & \textbf{97.5} & 97.9 & \textbf{99.8} & \textbf{96.4} & 96.5 & \textbf{4.45}\\
			\hline
		\end{tabular}
		\label{Compare}
	\end{center}
\end{table*}

\subsection{Comparison with State-of-the-Art Lightweight ECG Classification Models in Recent Years}
Table~\ref{Compare} compares the performance of the proposed model with state-of-the-art lightweight models from various perspectives.We have noticed a dearth of research on 17-class lightweight models for ECG. In order to ensure objective and fair comparison results, we have included Tuncer's full precision work\cite{b43} in Table~\ref{Compare}. 

Methodologically, our 1D model is easier to implement in hardware compared to Scrugi et al.'s\cite{b20} and Wong et al.'s\cite{b24} 2D  models.

In terms of bit-width, our model employs 1-bit values for both the activations and weights. This results in a storage footprint that is 8 times smaller than that of the model in work\cite{b20}, and 16-32 times smaller than that of the model in reference \cite{b21} when the number of weights are the same.

Regarding the number of distinguishable arrhythmia classes, our model is capable of recognizing 5 and 17 categories, whereas the works \cite{b21} and \cite{b24} are only able to recognize 2 categories.

In terms of evaluation metrics, we used 5 evaluation metrics, which made up for the shortcomings of insufficient evaluation metrics in work\cite{b23}. And the 5-class classification  model has the highest specificity of 98.5\% among 1-bit models, while the accuracy, specificity, and precision of  17-class classification model are among the highest in all 17-class models, with values of 97.5\%, 99.8\%, and 96.4\%, respectively.

In terms of data processing, our work only standardizes the data. This avoids the hardware considerations brought about by the complex data processing methods of works \cite{b20}, \cite{b21}, \cite{b23}, \cite{b43}. Compared to works \cite{b22} and \cite{b25}, which normalize the data, our accuracy is also 1.66\% and 1.23\% higher than theirs respectively.

Finally, our model's storage occupation is at the state-of-the-art level, with only 3.76 KB of storage required for the 5-class classification model and 4.45 KB for the 17-class classification model.

\subsection{Discussion}
In summary, by binarizing the weights and activations and fusing the operators into a threshold function, our arrhythmia 5-class classification model and 17-class classification model achieved a storage compression of 29.71 and 29.85 times, respectively, with only a slight decrease in accuracy of 0.19\% and 0.50\%. Even after transforming all the data in the dataset to $\left\{+1, -1\right\}$, our network still achieves an accuracy of no less than 91.50\%, which demonstrates its excellent anti-interference capability. Removing the fully connected layer and replacing the softmax layer further simplified the model's complexity and computational burden. While achieving the most advanced storage usage known, the accuracy was superior or equal to that of other models under the same classification standards. Additionally, the universal structure enables wearable devices to switch between 5-class and 17-class monitoring modes.

Due to the limitation of weights and activations to $\left\{+1, -1\right\}$ in the network, our BNN inevitably shows some gaps in evaluation metrics such as SEN and F1-score when compared to the full-precision network. In order to narrow the gap between the BNN and full-precision network in all aspects while maintaining the hardware friendliness of the BNN, this remains a challenging problem to be tackled in future research.

\section{Conclusion}
In this study, we propose a universal ultra lightweight BNN for 5-class and 17-class classification of long-term ECG signals. Computationally, all convolution operations in the model can be performed using XNOR gates, and the remaining main operations are mathematically fused into a threshold function, which almost completely eliminates multi-bit MAC and further reduces hardware energy consumption. Structurally, the regular structure facilitates future hardware implementation. In terms of model size, the memory consumed by storing the model is at the lowest known level. In the future, we will focus on the FPGA implementation of the BNN model to test its hardware acceleration effect and other performance parameters in real-life situations.

\def\bibfont{\footnotesize}


\begin{thebibliography}{00}
\bibitem{b1} World Health Organization, “Cardiovascular diseases (CVDs)” WHO Newsroom, 2021. Accessed: Feb. 27, 2023. [Online]. Available: https://www.who.int/news-room/fact-sheets/detail/cardiovasculardiseases-(cvds)
\bibitem{b2} J. J. Oresko et al., "A Wearable Smartphone-Based Platform for Real-Time Cardiovascular Disease Detection Via Electrocardiogram Processing," in IEEE Transactions on Information Technology in Biomedicine, vol. 14, no. 3, pp. 734-740, May. 2010.
\bibitem{b3} D. Wang et al., “Clinical Characteristics of 138 Hospitalized Patients With 2019 Novel Coronavirus–Infected Pneumonia in Wuhan, China,” JAMA, vol. 323, no. 11, p. 1061, Mar. 2020.
\bibitem{b4} A. M. Shaker, M. Tantawi, H. A. Shedeed and M. F. Tolba, "Generalization of Convolutional Neural Networks for ECG Classification Using Generative Adversarial Networks," in IEEE Access, vol. 8, pp. 35592-35605, 2020.
\bibitem{b5} A. Nasim, D. C. Nchekwube, F. Munir, and Y. S. Kim, “An Evolutionary-Neural Mechanism for Arrhythmia Classification With Optimum Features Using Single-Lead Electrocardiogram,” IEEE Access, vol. 10, pp. 99050–99065, 2022.
\bibitem{b6} M. Janveja, R. Parmar, M. Tantuway, and G. Trivedi, “A DNN-Based Low Power ECG Co-Processor Architecture to Classify Cardiac Arrhythmia for Wearable Devices,” IEEE Transactions on Circuits and Systems II: Express Briefs, vol. 69, no. 4, pp. 2281–2285, Apr. 2022.
\bibitem{b7} Y. Luo, K. -H. Teng, Y. Li, W. Mao, Y. Lian and C. -H. Heng, "A 74-$\mu$W 11-Mb/s Wireless Vital Signs Monitoring SoC for Three-Lead ECG, Respiration Rate, and Body Temperature," in IEEE Transactions on Biomedical Circuits and Systems, vol. 13, no. 5, pp. 907-917, Oct. 2019.
\bibitem{b8} Y. Zhao, Z. Shang and Y. Lian, "A 13.34 $\mu$W Event-Driven Patient-Specific ANN Cardiac Arrhythmia Classifier for Wearable ECG Sensors," in IEEE Transactions on Biomedical Circuits and Systems, vol. 14, no. 2, pp. 186-197, April. 2020.
\bibitem{b9} M. Courbariaux, I. Hubara, D. Soudry, R. El-Yaniv, and Y. Bengio, “Binarized Neural Networks: Training Deep Neural Networks with Weights and Activations Constrained to +1 or -1.” arXiv, Mar. 17, 2016. Accessed: Oct. 27, 2022. [Online]. Available: http://arxiv.org/abs/1602.02830
\bibitem{b10} A. Gotchev, N. Nikolaev and K. Egiazarian, "Improving the transform domain ECG denoising performance by applying interbeat and intra-beat decorrelating transforms," ISCAS 2001. The 2001 IEEE International Symposium on Circuits and Systems (Cat. No.01CH37196), Sydney, NSW, Australia, pp. 17-20,2001.
\bibitem{b11} L. Wang, W. Sun, Y. Chen, P. Li, and L. Zhao, “Wavelet Transform Based ECG Denoising Using Adaptive Thresholding,” in Proceedings of the 2018 7th International Conference on Bioinformatics and Biomedical Science, Shenzhen China, pp. 35–40, Jun. 2018.
\bibitem{b12} L. B. Marinho, N. de M. M. Nascimento, J. W. M. Souza, M. V. Gurgel, P. P. Rebouças Filho, and V. H. C. de Albuquerque, “A novel electrocardiogram feature extraction approach for cardiac arrhythmia classification,” Future Generation Computer Systems, vol. 97, pp. 564–577, Aug.2019.
\bibitem{b13} F. Jin, J. Liu, and W. Hou, “The application of pattern recognition technology in the diagnosis and analysis on the heart disease: Current status and future,” in 2012 24th Chinese Control and Decision Conference (CCDC), pp. 1304–1307, May. 2012.
\bibitem{b14} S. Parvaneh, J. Rubin, S. Babaeizadeh, and M. Xu-Wilson, “Cardiac arrhythmia detection using deep learning: A review,” Journal of Electrocardiology, vol. 57, pp. S70–S74, Nov. 2019.
\bibitem{b15} M. Nagel, M. Fournarakis, R. A. Amjad, Y. Bondarenko, M. van Baalen, and T. Blankevoort, “A White Paper on Neural Network Quantization.” arXiv, Jun. 15, 2021.
\bibitem{b16} W. Nogami, T. Ikegami, S. -i. O’uchi, R. Takano and T. Kudoh, "Optimizing Weight Value Quantization for CNN Inference," 2019 International Joint Conference on Neural Networks (IJCNN), Budapest, Hungary, pp. 1-8, 2019.
\bibitem{b17} Z. Li, H. Li, X. Fan, F. Chu, S. Lu, and H. Liu, “Arrhythmia Classifier Using a Layer-wise Quantized Convolutional Neural Network for Resource-Constrained Devices,” in Proceedings of the 2020 International Symposium on Artificial Intelligence in Medical Sciences, Beijing China, pp. 38–44, Sep. 2020.
\bibitem{b18} M. I. Rizqyawan, A. Munandar, M. F. Amri, R. Korio Utoro, and A. Pratondo, “Quantized Convolutional Neural Network toward Real-time Arrhythmia Detection in Edge Device,” in 2020 International Conference on Radar, Antenna, Microwave, Electronics, and Telecommunications (ICRAMET), pp. 234–239, Nov. 2020.
\bibitem{b19} M. M. Farag, ‘A Self-Contained STFT CNN for ECG Classification and Arrhythmia Detection at the Edge’, IEEE Access, vol. 10, pp. 94469–94486, 2022.
\bibitem{b20} M. A. Scrugli, D. Loi, L. Raffo, and P. Meloni, ‘An Adaptive Cognitive Sensor Node for ECG Monitoring in the Internet of Medical Things’, IEEE Access, vol. 10, pp. 1688–1705, 2022.
\bibitem{b21} G. Sivapalan, K. K. Nundy, S. Dev, B. Cardiff, and D. John, “ANNet: A Lightweight Neural Network for ECG Anomaly Detection in IoT Edge Sensors,” IEEE Transactions on Biomedical Circuits and Systems, vol. 16, no. 1, pp. 24–35, Feb. 2022.
\bibitem{b22} H. Sun et al., "Arrhythmia Classifier Using Convolutional Neural Network with Adaptive Loss-aware Multi-bit Networks Quantization," 2021 2nd International Conference on Artificial Intelligence and Computer Engineering (ICAICE), Hangzhou, China,pp. 461-467, 2021.
\bibitem{b23} Q. Wu, Y. Sun, H. Yan, and X. Wu, “ECG signal classification with binarized convolutional neural network,” Computers in Biology and Medicine, vol. 121, p. 103800, Jun. 2020.
\bibitem{b24} D. L. T. Wong, Y. Li, D. John, W. K. Ho, and C.-H. Heng, “An Energy Efficient ECG Ventricular Ectopic Beat Classifier Using Binarized CNN for Edge AI Devices,” IEEE Transactions on Biomedical Circuits and Systems, vol. 16, no. 2, pp. 222–232, Apr. 2022.
\bibitem{b25} A. Wang, W. Xu, H. Sun, N. Pu, Z. Liu, and H. Liu, “Arrhythmia Classifier using Binarized Convolutional Neural Network for Resource-Constrained Devices,” in 2022 4th International Conference on Communications, Information System and Computer Engineering (CISCE), Shenzhen, China, pp. 213–220, May 2022.
\bibitem{b26} M. Yun, S. Hong, S. Yoo, J. Kim, S.-M. Park, and Y. Lee, “Lightweight End-to-End Stress Recognition using Binarized CNN-LSTM Models,” in 2022 IEEE 4th International Conference on Artificial Intelligence Circuits and Systems (AICAS), pp. 270–273, Jun. 2022. 
\bibitem{b27} S. Aich and I. Stavness, “Global Sum Pooling: A Generalization Trick for Object Counting with Small Datasets of Large Images.” arXiv, Sep. 27, 2019.
\bibitem{b28} G. B. Moody and R. G. Mark, “The impact of the MIT-BIH Arrhythmia Database,” IEEE Eng. Med. Biol. Mag., vol. 20, no. 3, pp. 45–50, Jun. 2001.
\bibitem{b29} P. Pławiak, “Novel methodology of cardiac health recognition based on ECG signals and evolutionary-neural system,” Expert Systems with Applications, vol. 92, pp. 334–349, Feb. 2018.
\bibitem{b30} Testing and Reporting Performance Results of Cardiac Rhythm and STsegment Measurement Algorithms, ANSI/AAMI, AAMI, Association for the Advancement of Medical Instrumentation and American National Standards Institute, Arlington, VA, USA, 1999.
\bibitem{b33} W. Tang, G. Hua, and L. Wang, “How to Train a Compact Binary Neural Network with High Accuracy?,” AAAI, vol. 31, no. 1, Feb. 2017.
\bibitem{b38} R. Ding, H. Liu, and X. Zhou, “IE-Net: Information-Enhanced Binary Neural Networks for Accurate Classification,” Electronics, vol. 11, no. 6, p. 937, Mar. 2022, doi: 10.3390/electronics11060937.
\bibitem{b43} T. Tuncer, S. Dogan, P. Plawiak, and A. Subasi, “A novel Discrete Wavelet-Concatenated Mesh Tree and ternary chess pattern based ECG signal recognition method,” Biomedical Signal Processing and Control, vol. 72, p. 103331, Feb. 2022.
\end{thebibliography}
\end{document}